\documentclass[preprint,showpacs,preprintnumbers,amsmath,amssymb]{revtex4-1}
\usepackage{graphics,epsfig,subfigure}
\usepackage{color}

\begin{document}
\newcommand\beq{\begin{equation}}
\newcommand\eeq{\end{equation}}
\newcommand\beqn{\begin{eqnarray}}
\newcommand\eeqn{\end{eqnarray}}
\newcommand\nn{\nonumber}
\newcommand\fc{\frac}
\newcommand\lt{\left}
\newcommand\rt{\right}
\newcommand\pt{\partial}
\newcommand\tx{\text}


\title{Emergence of Cosmic Space and the Generalized Holographic Equipartition}
\author{Ke Yang\footnote{yangke09@lzu.edu.cn},
        Yu-Xiao Liu\footnote{liuyx@lzu.edu.cn, corresponding author},
        and Yong-Qiang Wang\footnote{yqwang@lzu.edu.cn}}.
 \affiliation{Institute of Theoretical Physics, Lanzhou University, Lanzhou 730000,
             China}

\begin{abstract}
Recently, a novel idea about our expanding Universe was proposed by T. Padmanabhan [arXiv:1206.4916]. He suggested that the expansion of our Universe can be thought of as the emergence of space as cosmic time progresses. The emergence is governed by the basic relation that the increase rate of Hubble volume is linearly determined by the difference between the number of degrees of freedom on the horizon surface and the one in the bulk. In this paper, following this idea, we generalize the basic relation to derive the Friedmann equations of an $(n+1)$-dimensional Friedmann-Robertson-Walker universe corresponding to general relativity, Gauss-Bonnet gravity, and Lovelock gravity.
\end{abstract}


\pacs{04.50.-h, 04.90.+e}



\maketitle


The tight connections between a gravitational system and a thermodynamic system, such as the four laws of black hole thermodynamics \cite{Bardeen1973}, motivate research on the idea that gravity may essentially be an emergent phenomenon.
In Ref. \cite{Sakharov1968}, Sakharov proposed the earliest version of this idea. He suggested that spacetime emerges as a mean field approximation of some underlying microscopic degrees of freedom (DOF). In Ref. \cite{Jacobson1995}, Jacobson derived the Einstein equations from the first law of thermodynamics on a local Rindler causal horizon.
In Ref. \cite{Verlinde2011}, Verlinde suggested that the gravity may not be a fundamental interaction but should be explained as an entropic force caused by changes of entropy associated with the information on the holographic screen. Further, with the holographic principle and the equipartition law of energy, the Newton's law of gravitation was derived. Moreover, a relativistic version of entropic force is able to give the Einstein equations. On the other hand, the Newton's law of gravitation was also derived by Padmanabhan in Ref. \cite{Padmanabhan2010a}, based on the equipartition law of energy and the relation between the entropy $S$ and active gravitational mass $E$, $S=E/(2T)$, with $T$ the Rindler temperature. In Ref. \cite{Li2012}, by combining the spirits of holographic and thermodynamic viewpoints, Li, Miao, and Meng succeeded in deriving all of the components of Einstein equations on a holographic timelike screen.
See also Refs. \cite{Visser2002,Padmanabhan2004,Barcelo2005,Hu2011} for examples and Refs. \cite{Padmanabhan2010,Padmanabhan2011} for a review.

However, most of these investigations just treat the gravitational field as an emergent phenomenon, but leave the spacetime as a preexisting background geometric manifold. Nevertheless, regarding the spacetime itself as an emergent structure may be a more complete way to view the emergence of gravitational phenomenon.
However, there are some conceptual difficulties associated with this idea. For example, it is very hard to think that the time used to describe the evolution of dynamical variables is emergent from some pregeometric variables and the space around finite gravitational systems is emergent.
Very recently, Padmanabhan proposed that these difficulties disappear when one considers the emergence of spacetime in cosmology, since the cosmic time of a geodesic observer provides a special choice of time variable, to which the observed  cosmic microwave background radiation is homogeneous and isotropic, and the spatial expansion of our Universe can be regarded as the consequence of the emergence of space. So he argued that \textit {the cosmic space can be emergent as cosmic time progresses} \cite{Padmanabhan2012}.

First, Padmanabhan noticed that the holographic principle is obeyed by a pure de Sitter universe with a Hubble constant $H$:
\beq
N_{\tx{sur}}=N_{\tx{bulk}}, \label{Holographic_Equipartition}
\eeq
where $N_{\tx{sur}}={4\pi H^{-2}}/{L^2_{\tx P}}$, with $L_{\tx{P}}$ the Planck length, is the number of DOF on the spherical surface of Hubble radius $H^{-1}$, and $N_{\tx{bulk}}={|E|}/[(1/2)T]$ with $|E|=|\rho+3p|V$, the Komar energy, is the effective number of DOF contained in this spherical volume $V=4\pi H^{-3}/3$ at the horizon temperature $T=H/2\pi$. Substituting the relation $\rho=-p$ for a de Sitter universe into Eq. (\ref{Holographic_Equipartition}), the standard result $H^2=8\pi L^2_{\tx{P}}\rho/3$ is recovered immediately.

Since condition (\ref{Holographic_Equipartition}) relates the DOF on the surface to the effective DOF in the bulk, Padmanabhan called it \textit{holographic equipartition}. However, when it deviates from the de Sitter case, such as our real Universe which is asymptotically de Sitter indicated by much astronomical evidence, the emergence of space occurs and the emergence must relate to the difference $\Delta N=N_{\tx{sur}}-N_{\tx{bulk}}$. And further, the emergence of space was considered to be equivalent to the expansion of our Universe. So he suggested that the basic law governing the emergence is simply \cite{Padmanabhan2012,Padmanabhan2012a}
\beq
\fc{dV}{dt}=L^2_{\tx{P}}\Delta N,
\label{Dynamical_Eq}
\eeq
where $t$ is the cosmic time and $V=4\pi H^{-3}/3$ the cosmic volume. After substituting $N_{\tx{sur}}={4\pi H^{-2}}/{L^2_{\tx P}}$ and $N_{\tx{bulk}}={|E|}/[(1/2)T]$, with $|E|$ the Komar energy $|E|=|\rho+3p|V$ and $T$ the horizon temperature $T=H/2\pi$, into the relation (\ref{Dynamical_Eq}), one gets the standard dynamical equation for a Friedmann-Robertson-Walker (FRW) universe in general relativity,
\beq
\fc{\ddot a}{a}=-\fc{4\pi L^{2}_{\tx{P}}}{3}(\rho+3p).
\label{FRW_Eq_GR}
\eeq
Moreover, making use of the continuity equation $\dot \rho+3H(\rho+p)=0$ and integrating Eq. (\ref{FRW_Eq_GR}), one gets the Friedmann equation
\beq
H^2+\fc{k}{a^2}=\fc{8\pi L^2_{\tx{P}}}{3}\rho,
\eeq
where $k$ is an integral constant and can be related to the spatial curvature of the FRW universe.

In Ref.~\cite{Cai2012b}, Cai generalized the derivation process to the higher $(n+1)$-dimensional spacetime. By taking account of the entropy formulas of black holes modified in Gauss-Bonnet gravity and Lovelock gravity, the number of DOF on the holographic surface $N_{\tx{sur}}$ and the volume increase were modified, until finally, the Friedmann equations of a higher-dimensional FRW universe in these gravity theories were obtained.

Inspired by the work of Cai~\cite{Cai2012b}, we would like to give another generalization of the work of Padmanabhan \cite{Padmanabhan2012}. Our generalization is based on the following simple concepts:
(a) The number of DOF on the surface of the Hubble sphere $N_{\tx{sur}}$ accounts for the number of bits stored on the surface, and it is proportional to the area of the surface regardless of the gravity theories; the bulk DOF in the Hubble sphere $N_{\tx{bulk}}$ obeys the equipartition law of energy.
(b) The emergence of cosmic space generally relates to both the number of DOF on the holographic surface $N_{\tx{sur}}$ and the difference $\Delta N$ between the surface and bulk (or equivalently relates to both  $N_{\tx{bulk}}$ and $\Delta N$, or both $N_{\tx{sur}}$ and $N_{\tx{bulk}}$).
(c) The different ways of the emergence of cosmic space result in different dynamical equations which can be related to the different gravity theories and can be verified by astronomical observations.
Thus we prefer to generalize the basic dynamical equation (\ref{Dynamical_Eq}) in an $(n+1)$-dimensional universe instead of modifying the magnitudes such as $N_{\tx{sur}}$ and so on. Now the general form of a dynamical equation is proposed as
\beq
\fc{dV}{dt}=L^{n-1}_{\tx{P}}f(\Delta N, N_{\tx{sur}}),
\label{Modified_Dynamical_Eq}
\eeq
where the function $f(\Delta N, N_{\tx{sur}})$ determines the evolution of our Universe. In an $(n+1)$-dimensional universe, the number of DOF on the holographic surface is given by \cite{Cai2012b}
\beq
N_{\tx{sur}}={\alpha A}/{L^{n-1}_{\tx{P}}}, \label{Nsur}
\eeq
where
\beq
A=n\Omega_n/H^{n-1}, \label{Area}
\eeq
with $\Omega_n$ the volume of an $n$ sphere of unit radius and $\alpha=(n-1)/2(n-2)$, and the number of DOF in the spherical volume
\beq
V=\Omega_n/H^n \label{volume}
\eeq
is given by
\beq
N_{\tx{bulk}}=\frac{|E|}{T/2}, \label{Nbulk}
\eeq
where $E=\frac{(n-2)\rho+np}{n-2}V$ is the bulk Komar energy \cite{Cai2010} and $T=H/2\pi$ the temperature of the Hubble horizon. As in Refs. \cite{Padmanabhan2012,Cai2012b}, here we only consider the accelerating phase with $(n-2)\rho+np<0$.

First, we choose the function $f(\Delta N, N_{\tx{sur}})$ as the most simplest form \cite{Padmanabhan2012,Cai2012b}
\beq
f(\Delta N)={\Delta N}/{\alpha}.
\label{Modified_f_GR}
\eeq
It means that the emergence of space relates only to the difference of the number of DOF between the surface and the bulk. Equation (\ref{Modified_Dynamical_Eq}) just reduces to the relation (\ref{Dynamical_Eq}) for $n=3$. Now, with the above expressions for $N_{\tx{sur}}$, $N_{\tx{bulk}}$, and $V$, the general relation (\ref{Modified_Dynamical_Eq}) just gives
\beq
\fc{\ddot a}{a}=-\fc{8\pi L^{n-1}_{\tx{p}}}{n(n-1)}[(n-2)\rho+np].
\eeq
Further, utilizing the $(n+1)$-dimensional continuity equation $\dot \rho+nH(\rho+p)=0$, one finally arrives at the standard Friedmann equation of the $(n+1)$-dimensional FRW universe in general relativity:
\beq
H^2+\fc{k}{a^2}=\fc{16\pi L^{n-1}_{\tx{p}}}{n(n-1)}\rho,
\label{Friedmann_Eq_GR}
\eeq
where $k$ is an integral constant and relates to the spatial curvature.

Next, suppose that the emergence of space relies on a more complicated relation about the number of DOF on the surface $N_{\tx{sur}}$ and the number of DOF in the bulk $N_{\tx{bulk}}$. Here it is chosen as the form
\beq
f(\Delta N, N_{\tx{sur}})=\fc{{\Delta N}/{\alpha}+\tilde\alpha K\lt({N_{\tx{sur}}}/{\alpha}\rt)^{1+\fc{2}{1-n}}}{1+2\tilde\alpha K\lt({N_{\tx{sur}}}/{\alpha}\rt)^{\fc{2}{1-n}}},
\label{Modified_f_GB}
\eeq
where $K=({n\Omega_n}/{L^{n-1}_{\tx{P}}})^{\fc{2}{n-1}}$.
The coefficient $\tilde \alpha$ is a parameter with the length dimension 2, and it can be identified as something we are familiar with from the subsequent dynamical equation. When $\tilde \alpha=0$, this choice (\ref{Modified_f_GB}) reduces to the previous one (\ref{Modified_f_GR}). And further, making use of the expressions of $N_{\tx{sur}}$, $N_{\tx{bulk}}$, and $V$, after some simple algebra, Eq. (\ref{Modified_Dynamical_Eq}) gives
\beq
(1+2\tilde \alpha H^2)\dot H+(1+\tilde \alpha H^2)H^2=-\fc{8\pi L^{n-1}_{\tx{P}}}{n(n-1)}[(n-2)\rho+np].
\eeq
Then, with the $(n+1)$-dimensional continuity equation and the above equation, we finally get the following dynamical equation:
\beq
H^2+\tilde \alpha H^4=\fc{16\pi L^{n-1}_{\tx{P}}}{n(n-1)}\rho.
\label{Friedmann_Eq_GB}
\eeq
Here the integral constant is set to zero. This dynamical equation is nothing but the standard Friedmann equation of the $(n+1)$-dimensional spatial flat FRW universe in Gauss-Bonnet gravity, and hence, the parameter $\tilde \alpha$ can just be regarded as the Gauss-Bonnet coefficient \cite{Cai2005}.

Further, we suppose the function $f(\Delta N, N_{\tx{sur}})$ is a more general form than the Gauss-Bonnet case, i.e.,
\beq
f(\Delta N, N_{\tx{sur}})=\fc{{\Delta N}/{\alpha}+\sum^{m}_{i=2} \tilde c_i K_i\lt({N_{\tx{sur}}}/{\alpha}\rt)^{1+\fc{2i-2}{1-n}}}{1+\sum^{m}_{i=2}i\tilde c_i K_i\lt({N_{\tx{sur}}}/{\alpha}\rt)^{\fc{2i-2}{1-n}}},
\label{Modified_f_Lovelock}
\eeq
where $K_i=(n\Omega_n/{L^{n-1}_{\tx{P}}})^{\fc{2i-2}{n-1}}$, $m=[n/2]$, and $\tilde c_i$ are some coefficients with the length dimension $(2i-2)$, and especially $\tilde c_1=1$. If $\tilde c_i=0 $ for $i>2$, then this assumption recovers the Gauss-Bonnet one. So with the same derivation process of the former case, one obtains the dynamical equation
\beq
\lt(\sum^{m}_{i=1}i\tilde c_i H^{2i-2}\rt) \dot H+\sum^{m}_{i=1}\tilde c_i H^{2i}=-\fc{8\pi L^{n-1}_{\tx{P}}}{n(n-1)}[(n-2)\rho+np].
\eeq
Making use of the $(n+1)$-dimensional continuity equation again and after integrating, one finally arrives at the equation
\beq
\sum^{m}_{i=1}\tilde c_i H^{2i}=\fc{16\pi L^{n-1}_{\tx{p}}}{n(n-1)}\rho.
\label{Friedmann_Eq_Lovelock}
\eeq
Here the integration constant is also set to zero. Equation (\ref{Friedmann_Eq_Lovelock}) is just the standard Friedmann equation of the $(n+1)$-dimensional spatial flat FRW universe in Lovelock gravity and these parameters $\tilde c_i$ could be regarded as the coefficients in front of Euler densities $\mathcal{L}_i$ \cite{Cai2005}.

Here we note that the Friedmann equations (\ref{Friedmann_Eq_GR}), (\ref{Friedmann_Eq_GB}), and (\ref{Friedmann_Eq_Lovelock}) were also obtained by Cai in \cite{Cai2012b}, where he modified the number of DOF on the surface of the Hubble sphere and introduced the increase of an effective volume in Gauss-Bonnet gravity and Lovelock gravity. However, instead of modifying the number of DOF on the surface of the Hubble sphere and the volume increase, we derive these equations from the generalized law governing the emergence of cosmic space (\ref{Modified_Dynamical_Eq}). Since the evolution of our Universe, which is described by Friedmann equations, can be regarded as the emergence of comic space, different functions $f(\Delta N, N_{\tx{sur}})$ could be interpreted as Friedmann equations corresponding to different gravity theories.
It is interesting that the simplest emergent way that only associates with the difference $\Delta N$ of the numbers of DOF between surface and bulk corresponds to general relativity. Here, our method has the same problem as presented in \cite{Cai2012b}; i.e., we also cannot derive the Friedmann equations of the spatial nonflat FRW universe corresponding to the Gauss-Bonnet gravity and the Lovelock gravity.


In de Sitter universe, $H$ is a constant; thus, $dV/dt=0$ and $f(\Delta N, N_{\tx{sur}})=0$. Then from the functions $f(\Delta N, N_{\tx{sur}})$ given in Eqs.~(\ref{Modified_f_GR}), (\ref{Modified_f_GB}), and (\ref{Modified_f_Lovelock}), it is clear that the difference $\Delta N$ is generally nonvanishing. For example, ${\Delta N}=-{\alpha}\tilde\alpha K\lt({N_{\tx{sur}}}/{\alpha}\rt)^{1+\fc{2}{1-n}}$ for Gauss-Bonnet gravity. Thus the holographic equipartition $\Delta N=0$ (which is equivalent to $|E|=\frac{1}{2}N_{\tx{sur}}T$ for general relativity) should be correspondingly generalized to $f(\Delta N, N_{\tx{sur}})=0$. Moreover,  for the case of Gauss-Bonnet gravity, the standard equipartition law $|E|=\frac{1}{2}N_{\tx{sur}}T$ hold in general relativity
should be correspondingly changed to $|E|=\frac{1}{2}N_{\tx{sur}}T\lt[1+\tilde\alpha K\lt({N_{\tx{sur}}}/{\alpha}\rt)^{\fc{2}{1-n}}\rt]$.

\acknowledgments{
The authors would like to thank Jie Yang and Shao-Wen Wei for helpful discussions.
This work was supported in part by the Program for New Century Excellent Talents
in University, the National Natural Science Foundation of China (Grant No. 11075065),
the Doctoral Program Foundation of Institutions of Higher Education of China
(Grant No. 20090211110028), and the Huo Ying-Dong Education Foundation of
Chinese Ministry of Education (Grant No. 121106).}




\end{document}